\documentclass[12pt]{article}
\usepackage{graphicx}
\begin{document}
\bigskip
\centerline{\Large \bf Positive mutations and mutation-dependent}

\medskip
\centerline{\Large \bf Verhulst factor in Penna ageing model}
\bigskip

S. Moss de Oliveira$^1$, D. Stauffer$^2$, P.M.C de Oliveira$^1$ 
and J.S. S\'a Martins$^1$.

\medskip
Laboratoire PMMH, \'Ecole Sup\'erieure de Physique et de Chimie
Industrielles, 10 rue Vauquelin, F-75231 Paris, Euroland

\medskip

\noindent
$^1$ Visiting from Instituto de F\'{\i}sica, Universidade
Federal Fluminense; Av. Litor\^{a}nea s/n, Boa Viagem,
Niter\'{o}i 24210-340, RJ, Brazil; suzana@if.uff.br, jssm@if.uff.br, 
pmco@if.uff.br.
\medskip

\noindent
$^2$ Visiting from Institute for Theoretical Physics, Cologne
University, D-50923 K\"oln, Euroland; stauffer@thp.uni-koeln.de

\bigskip

Abstract: We twice modify the Penna model for biological ageing. First 
we introduce back (good) mutations and a memory for them into the model. It 
allows us to observe an improvement of the species fitness over long time 
scales as well as punctuated equilibrium. Second we adopt a food/space 
competition factor that depends on the 
number of accumulated mutations in the individuals genomes, and get rid of 
the fixed limiting number of allowed mutations. Besides reproducing the main 
results of the standard model, we also observe a mortality maximum 
for the oldest old.  
  
\medskip



\section{Introduction: the standard Penna model}

For biological ageing, the Penna model \cite{penna} presently
is the most widespread computer simulation method. The genome
of each individual is given by a string of 32 bits,
representing dangerous inherited diseases (detrimental
mutations) for the at most 32 time intervals during the life of this
individual. A 0-bit means health, a 1-bit on position $a$ of
the bit-string means a harmful mutation affecting the health from that
age $a$ on. Three such diseases kill the individual at that
age $a$ at which the third disease becomes active.
Besides these deaths from genetic reasons, individuals also die 
at each time step with the Verhulst probability $N/N_{max}$ 
where $N$ is the total population and $N_{max}$ a constant parameter, often
called the carrying capacity. At each
time step, i.e. one iteration of the whole population,
each living individual above a
minimum reproduction age of 8 gives birth to children
with the same genome as the mother except for one mutation in 
a randomly selected position. For more results from the Penna
model we refer to \cite{book,stauffer}. In particular, this
model was used to compare various forms of sexual and asexual
reproduction for diploids \cite{anais}.

\section{Model with good mutations}

We have been often criticized for working only with bad mutations. In 
this case the bit position randomly selected in the parent's genome has  
its bit set to one independently of its previous value. So the 
offspring genome is worse or equal to that of the parent, but never better. 
How can it then be compatible with the Darwinian evolution of more complex 
and fitter species? The answer is that the Penna model was invented originally 
\cite{penna} to describe ageing, which happens for {\it homo sapiens}
within a century, and not to describe the evolution of 
{\it homo} over millions of years after the
separation from other primates. On short time scales, nearly all mutations are
bad; on long time scales, the rare good mutations dominate Darwinian evolution
of fitter species. We now aim to include this effect of two time scales into
the Penna model. 
 
We start with the asexual Fortran program published in \cite{book} with one
bit-string of 32 bits as standard ``genome'', and with one child born per 
surviving adult parent per iteration. When the child is born, a randomly 
selected bit of the standard genome becomes set (one) if it was zero before; it
stays set otherwise. When in a first modification we allowed positive mutations
(reversing a set bit to a zero bit) with probability 0.01, we found improved 
fitness as shown by a higher population, but no evolution over long times.

In our second modification, we return to the case where only bad mutations 
are allowed: genetic improvement will be represented now by the possibility 
of not counting some of them, as follows. Each individual gets a second 
such bit-string as memory for good mutations. At birth, this memory genome
gets, with a low probability $q$, on that same bit position where the standard 
genome is mutated, a bit set to one if it was zero before (otherwise it stays
at one). In the evaluation of the active number of
mutations (three of which are lethal) those bit positions 
are ignored. That is, if an individual has a bit set in a given position of the 
standard genome and also a bit set in the same position of the memory genome, 
then that mutation is ignored (similar
to recessive mutations in sexual reproduction \cite{book,anais}). 
Now we found the desired long-term evolution,
but it approached the unrealistic ideal of no active mutations at all, which 
also means no ageing. The reason is that if a bit is set in the memory string 
of the parent, it stays necessarily set in the offspring one. 
\medskip

Only our third modification had the desired effect of allowing long-term 
improvement but still with ageing. At birth, the above memory genome gets,
again on the same position as the mutation in the standard genome, a bit set to
zero with probability $q$; if it was already zero it stays at zero. Thereafter, 
with an even smaller probability $q^2$, the same bit is set to one. Then,
the two mutated bit-strings, i.e. the standard and the memory genome,
are given on to the offspring. Only the standard mutations at positions 
where the memory bit is zero have detrimental effects. 
The mutation part of the modified program now is:

\begin{verbatim}            
c             select a random position to mutate  
c             in the standard genome 
              ibm=ibm*16807
              p=bit(ishft(ibm,-59))
              gene1=ior(gene1,p)
              if(bad) goto 13
c             mutation at same position of the 
c             memory bit-string 
              ibm=ibm*mult
              if(ibm.lt.iprob) gen2f(i)= ior(gen2f(i),p)
              if(ibm.lt.ipro2) gen2f(i)=ieor(gen2f(i),p)
 13           continue
\end{verbatim}

Here, {\tt ipro2} corresponds to the probability $q$, {\tt iprob} to the 
probability $q(1+q)$, {\tt gene1} to the standard genome, and {\tt gen2f(i)} 
to the new memory bit-string introduced here. Note that the same random
bit {\tt p} controls the mutations in the two bit-strings, the array 
${\rm bit(i) = 2^i}$ previously defined. (Our random
integers {\tt ibm} used 64 bits and multiplication with {\tt mult} = 
$13^{13}$. The logical variable {\tt bad}   
is true only during the first 10 percent of the simulation).
 
\begin{figure}[!ht]
\begin{center}
\includegraphics[angle=-90,scale=0.5]{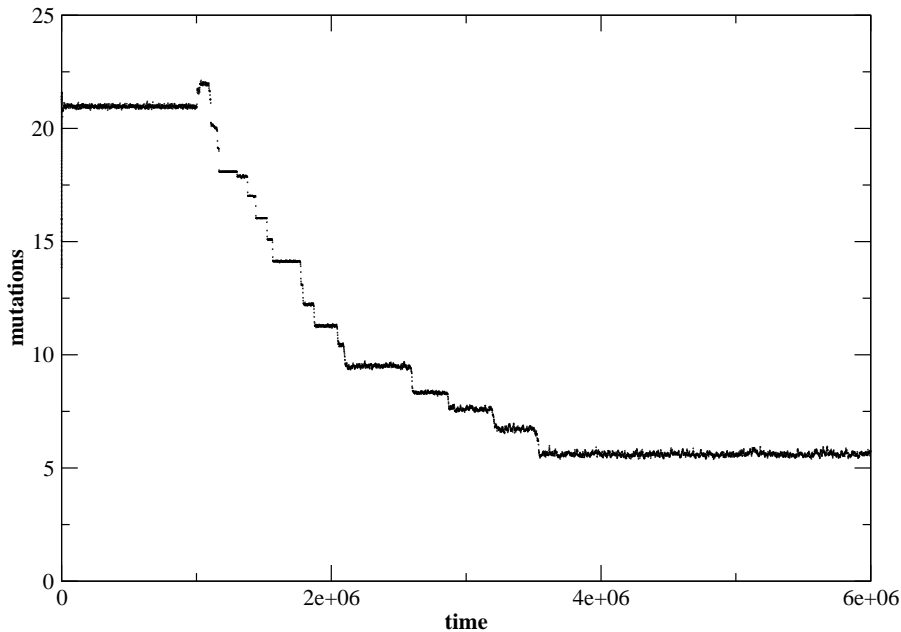}
\end{center}
\caption{Darwinian evolution of a fitter species over long times
at low probability $q = 0.001$ with one child per birth; carrying capacity
200,000; actual population increases from about 26,000 to about 37,000 when
good mutations on the memory bit-string are switched on at time $10^6$.
}
\end{figure}

Figure 1 shows for $q=0.001$ the effects of the new memory bit-string for good 
mutations. First we equilibrate the standard model, starting from a
random standard genome where on average 16 of the 32 bits are set, and a
completely empty memory genome. After one million iterations, the mutations
in the memory genome are switched on, allowing for good mutations with 
probability $q^2 = 10^{-6}$. From now on the average number of bad mutations
in the whole standard bit-string (not counting those made recessive by the 
memory bit-string) goes down to a much lower value (from 16 to 5). The resulting
higher fitness increases the population (not shown). We also see clearly that
the improvement of the species proceeds mostly in rapid steps followed by
longer intervals of constant number of mutations: ``Punctuated equilibrium'' as 
in reality \cite{punct}. With $q = 0.1$ or 0.01, this improvement proceeds 
faster (not shown). 

Thus, the Penna model allows also an improvement of fitness over long time 
scales.

\section{Model with mutation-dependent Verhulst factor}

As out pointed in section 1, in the standard Penna model there is a random time-dependent 
killing factor $V = N(t)/N_{max}$, known as the Verhulst factor. 
At each iteration and for each individual $i$   
a random number $r_i$ between zero and one is generated and compared with $V$: if 
$r_i \le V$, the individual dies independently of its age or genome. Its main 
purpose is to avoid the exponential increase of the population through a competion 
for food and space inside an environment that can carry at most $N_{max}$ 
individuals. However, in real populations the fitter the animal is, the higher is 
its probability to win a dispute. So this random killing factor is frequently  
criticized in the literature \cite{verhulst}. 

We now introduce a new Verhulst that mostly depends on each individual's current number 
of mutations, ${\rm mut}$, active at it's age: 
$$V_i = \frac {N(t)} {N_{max} \times f_i({\rm mut})} \, , \,\,\,\,\, {\rm with} \,\,\, 
f(0)=1 \,\,\,\, {\rm and} \,\,\,f({\rm mut}) = \frac {1} {n^{\rm mut}}\,\, .$$ 

Now if an individual has not yet accumulated any bad mutation at a given age, 
it still can be killed by the standard Verhulst probability at that 
age (since $f(0)=1$). However, whenever a new mutation appears, the carrying capacity for 
that individual is divided by some fixed constant $n$, which increases it's probability 
to die by the 
same factor. Also we don't consider anymore a limiting number of bad mutations: 
the individual dies either by this new Verhulst or because it reaches age 32.    
\medskip

\begin{figure}[!ht]
\begin{center}
\includegraphics[angle=-90,scale=0.5]{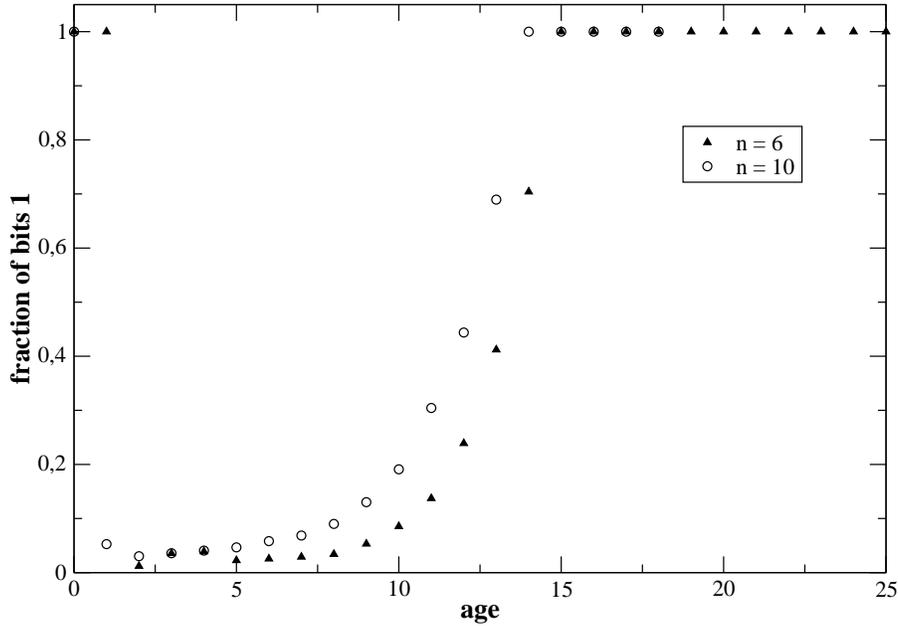}
\end{center}
\caption{Average fraction of bits set to one per age. Parameters: average population 
around 120,000 individuals; carrying capacity = 1 million; one child per birth; 
minimum reproduction age = 8; averages taken over the last 10,000 time steps of 
a total of 1 million time steps. Circles correspond to $n=10$ in the Verhulst factor 
and triangles to $n=6$.   
}
\end{figure}

Figure 2 shows the relative number of individuals with a bit set at a given age,  
for $n=6$ and for $n=10$. We see that before the minimum reproduction age $R=8$, 
for $n=6$ all 
individuals have bits set at ages zero and one, while for $n=10$ the fixation 
occurs only at age one. 
In fact, the smaller the value of $n$ is, the higher is the number of fixed bits before 
$R$. The shape of these curves is similar to those obtained with the standard 
model \cite{whysex}, although in that case the asexual populations always fix $T-1$ 
bits before $R$, where $T$ is the limit number of allowed diseases. For this reason 
its longevities are smaller than in the present case, where there is no 
threshold for bad mutations. 

\medskip

\begin{figure}[!h]
\begin{center}
\includegraphics[angle=-90,scale=0.5]{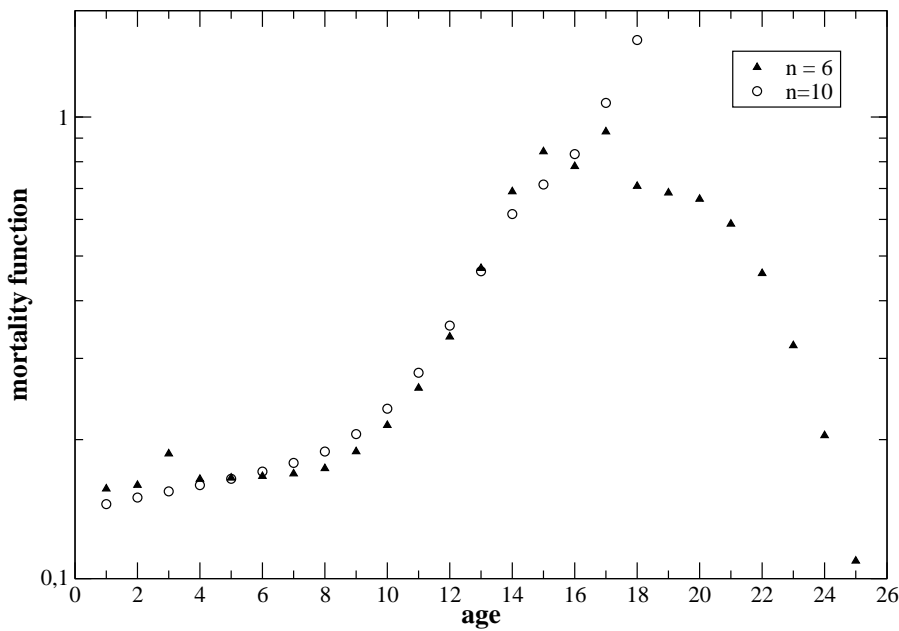}
\end{center}
\caption{Mortality functions (semi-logarithmic scale) for both cases; same parameters 
and symbols as before.
}
\end{figure}

Finally, figure 3 shows the mortality functions, $q(a)$, where $D_a$ is the number of 
deaths at age $a$: $$q(a) = - \ln \left [1 - \frac {D_a} {N_a} \right ] \, \, .$$ 
Again for the $n=10$ case, the mortality is roughly the same as the usual one, 
increasing exponentially with increasing age, for ages above the minimum reproduction 
age (here equal to 8). Such behaviour is known as the Gompertz low of mortality 
\cite{gompertz}. For $n=6$, the result changes dramatically and the mortality 
decelerates at older ages, as observed for
drosophilas \cite{curtsinger,promislow}, and may even decrease, as observed for
medflies \cite{carey}. (For a review on experimental results
see \cite{zeus}). A mortality plateau was also shown to occur 
within the analytical solution of a slightly modified version of the Penna model \cite{prl} 
where the strict death rule for a fixed number of accumulated mutations is relaxed, 
although not removed. It was also obtained 
before with the standard Penna model, but using more complicated strategies as for instance,  
assuming a rather unusual antagonistic pleiotropy \cite{pleio}. Alternative simulations by 
biologists thus far failed to get this mortality maximum \cite{biol}.
\medskip

It is important to say that in the figures presented here we have considered 
all the results except those for which there was, on average, less than one alive 
individual with a given age. For this reason, the statistics concerning extremely 
old individuals is poor. However, such cases correspond to the tails of the curves    
and don't perturb the reliability of our mortality maximum.   
For instance, in the $n=6$ case, there is, on average, only one alive individual   
with age 25, but there are around 150 with age 18 and 730 with age 16, where the 
effect appears. 

\section{Conclusions}

Introducing into the Penna bit-string model good mutations or a memory to discount 
the bad ones (which roughly plays the role of changing a dominant mutation into a 
recessive one), we obtain an improvement of the population fitness over long time 
scales. This improvement proceeds mostly in rapid steps followed by longer 
intervals of constant fitness, meaning that punctuated 
equilibrium can also be obtained within the Penna model.

We also modify the original model changing the completely random killing Verhulst 
factor into another one that increases exponentially with the number of active mutations.
In this case we don't consider anymore that a given limiting number of mutations 
kills the individual. Now some individuals can live longer, despite carrying a 
higher number of genetic diseases, than others which inherited a smaller one. 
With this modification we reproduce the main features of the original model and 
also obtain a mortality maximum at old ages.    
\medskip

\noindent {\bf Acknowledgements}: To PMMH at ESPCI for the warm hospitality,
to Sorin T\u{a}nase-Nicola for helping us with the computer facilities;
SMO, PMCO and JSSM thank the Brazilian agencies FAPERJ and CNPq for financial 
support.

\end{document}